\documentclass[11pt]{article}
\usepackage{amssymb,amsmath,amsfonts}
\usepackage{graphicx}
\usepackage{graphics}
\usepackage{eepic,epsfig}

\begin{document}

\title{Summation formula over the zeros of the associated Legendre function
with a physical application}
\author{A. A. Saharian\thanks{%
E-mail: saharian@ictp.it } \\
\textit{Department of Physics, Yerevan State University, }\\
\textit{1 Alex Manogian Street, 0025 Yerevan, Armenia }\\
\textit{and}\\
\textit{Departamento de F\'{\i}sica, Universidade Federal da Para\'{\i}ba,}\\
\textit{\ 58.059-970, Caixa Postal 5.008, Jo\~{a}o Pessoa, PB, Brazil}}
\maketitle

\begin{abstract}
Associated Legendre functions arise in many problems of mathematical
physics. By using the generalized Abel-Plana formula, in this paper we
derive a summation formula for the series over the zeros of the associated
Legendre function of the first kind with respect to the degree. The
summation formula for the series over the zeros of the Bessel function,
previously discussed in the literature, is obtained as a limiting case. The
Wightman function for a scalar field with general curvature coupling
parameter is considered inside a spherical boundary on background of
constant negative curvature space. The corresponding mode sum contains
series over the zeros of the associated Legendre function. The application
of the summation formula allows us to present the Wightman function in the
form of the sum of two integrals. The first one corresponds to the Wightman
function for the bulk geometry without boundaries and the second one is
induced by the presence of the spherical shell. For points away from the
boundary the latter is finite in the coincidence limit. In this way the
renormalization of the vacuum expectation value of the field squared is
reduced to that for the boundary-free part.
\end{abstract}

\bigskip


\bigskip

\section{Introduction}

\label{sec:Introd}

In a number of problems in mathematical physics we need to sum over the
values of a certain function at integer points, and then subtract the
corresponding integral. In particular, in quantum field theory the
expectation values for physical observables induced by the presence of
boundaries are presented in the form of this difference. The corresponding
sum and integral, taken separately, diverge and some physically motivated
procedure to handle the finite result, is needed. For a number of boundary
geometries one of the most convenient methods to obtain such renormalized
values is based on the use of the Abel-Plana summation formula \cite%
{Hard91,Henr74} (for different forms of this formula discussed in the
literature see also \cite{Saha07Rev}). Applications of the Abel-Plana
formula in physical problems related to the Casimir effect for flat boundary
geometries and topologically non-trivial spaces with corresponding
references can be found in \cite{Grib94,Most97}. The use of this formula
allows to extract in a cutoff independent way the Minkowski vacuum part and
to obtain for the renormalized part rapidly convergent integrals useful, in
particular, for numerical calculations.

However, the applications of the Abel-Plana formula in its standard form are
restricted to the problems where the normal modes are explicitly known. In
\cite{Sah1} we have considered a generalization of this formula, which
essentially enlarges the application range and allows to include problems
where the eigenmodes are given implicitly as zeros of a given function. Well
known examples of this kind are the boundary-value problems with spherical
and cylindrical boundaries. The generalized Abel-Plana formula contains two
meromorphic functions and by specifying one of them the Abel-Plana formula
is obtained (for other generalizations of the Abel-Plana formula see \cite%
{Most97,Bart80,Zaya88}). Applying the generalized formula to Bessel
functions, in \cite{Sah1, Sahdis} summation formulae are obtained for the
series over the zeros of various combinations of these functions (for a
review with physical applications see also Ref. \cite%
{Saha07Rev,Saha00Rev,Saha06PoS}).

The summation formulae derived from the generalized Abel-Plana formula have
been applied for the evaluation of the vacuum expectation values of local
physical observables in the Casimir effect (for the Casimir effect see \cite%
{Grib94,Most97,Plun86}) for plane boundaries with Robin or non-local
boundary conditions \cite{Rome02}, for spherical boundaries in Minkowski and
global monopole bulks \cite{Saha01} and for cylindrical boundaries in
Minkowski and cosmic string bulks \cite{Rome01}. By making use of the
generalized Abel-Plana formula, the vacuum expectation values of the field
squared and the energy-momentum tensor in closely related but more
complicated geometry of a wedge with cylindrical boundary are investigated
in \cite{Reza02} for both scalar and electromagnetic fields. As in the case
of the Abel-Plana formula, the use of the generalized formula in these
problems allows to extract the contribution of the unbounded space and to
present the boundary-induced parts in terms of exponentially converging
integrals. In \cite{SahaRind1} summation formulae for the series over the
zeroes of the modified Bessel functions with an imaginary order are derived
by using the generalized Abel-Plana formula. This type of series arise in
the evaluation of the vacuum expectation values induced by plane boundaries
uniformly accelerated through the Fulling-Rindler vacuum. Another class of
problems where the application of the generalized Abel-Plana formula
provides an efficient way for the evaluation of the vacuum expectation
values is considered in \cite{Saha05b}. In these papers braneworld models
with two parallel branes on anti-de Sitter bulk are discussed. The
corresponding mode-sums for physical observables bilinear in the field
contain series over the zeroes of cylinder functions which are summarized by
using the generalized Abel-Plana formula. The geometry of spherical branes
in Rindler-like spacetimes is considered in \cite{Saha07RindBr}. In \cite%
{Saha07Helic} from the generalized Abel-Plana formula a summation formula is
derived over the eigenmodes of a dielectric cylinder and this formula is
applied for the evaluation of the radiation intensity from a point charge
orbiting along a helical trajectory inside the cylinder.

The physical importance of the Bessel functions is related to the fact that
they appear as solutions of the field theory equations in various
situations. In particular, in spherical and cylindrical coordinates the
radial parts of the solutions for the scalar, fermionic, and electromagnetic
wave equations on background of the Minkowski spacetime are expressed in
terms of these functions. Another important class of special functions is
the so-called Legendre associated functions (see, for instance, \cite%
{Erde53a,Abra72}). These functions can be considered as
generalizations of the Bessel functions: in the limit of large
values of the degree when the argument is close to unity they
reduce to the Bessel functions. The associated Legendre functions
arise naturally in many mathematical and physical applications. In
particular, they appear as solutions of physical field equations
on background of constant curvature spaces (see, for instance,
\cite{Grib94,Most97,Birr82}) and the above-mentioned limit
corresponds to the limit when the curvature radius of the bulk
goes to infinity. The eigenfunctions in braneworld models with de
Sitter and anti-de Sitter branes are also expressed in terms of
the Legendre functions (see \cite{Noji00}). Motivated by this, in
the present paper, by making use of the generalized Abel-Plana
formula, we obtain a summation formula for the series over the
zeros of the associated Legendre function of the first kind with
respect to the degree. In particular, this type of series appear
in the evaluation of expectation values for physical observables
bilinear in the operator of a quantum field on background of
constant curvature spaces in the presence of boundaries. As in the
case of the other Abel-Plana-type formulae, previously considered
in the literature, the formula discussed here presents the sum of
the series over the zeros of the associated Legendre function in
the form of the sum of two integrals. In boundary-value problems
the first one corresponds to the situation when the boundary is
absent and the second one presents the part induced by the
boundary. For a large class of functions the latter is rapidly
convergent and, in particular, is useful for the numerical
evaluations of the corresponding physical characteristics.

We have organized the paper as follows. In the next section, by specifying
the functions in the generalized Abel-Plana formula we derive a formula for
the summation of series over zeros of the associated Legendre function with
respect to the degree. In section \ref{sec:Special}, special cases of this
summation formula are considered. First, as a partial check we show that as
a special case the standard Abel-Plana formula is obtained. Then we show
that from the summation formula discussed in section \ref{sec:SumForm}, as a
limiting case the formula is obtained for the summation of the series over
the zeros of the Bessel function, previously derived in \cite{Sah1}. A
physical application is given in section \ref{sec:Phys}, where the positive
frequency Wightman function for a scalar field is evaluated inside a
spherical boundary on background of a negative constant curvature space. It
is assumed that the field obeys Dirichlet boundary condition on the
spherical shell. The use of the summation formula from section \ref%
{sec:SumForm} allows us to extract from the vacuum expectation value the
part corresponding to the geometry without boundaries and to present the
part induced by the spherical shell in terms of an integral, which is
rapidly convergent in the coincidence limit for points away from the
boundary. The main results of the paper are summarized in section \ref%
{sec:Conclus}. In appendix \ref{sec:Zeros} we show that the zeros of the
associated Legendre function of the first kind with respect to the degree
are simple and real, and the asymptotic form for large zeros is discussed.
In appendix \ref{sec:LegAsymp} asymptotic formulae for the associated
Legendre functions are considered for large values of the degree. These
formulae are used in section \ref{sec:SumForm} to obtain the constraints
imposed on the function appearing in the summation formula.

\section{Summation formula}

\label{sec:SumForm}

In this section we derive a summation formula for the series over zeros of
the associated Legendre function of the first kind, $P_{iz-1/2}^{\mu }(u)$,
with respect to the degree, assuming that $u>1$ and $\mu \leqslant 0$ (in
this paper the definition of the associated Legendre functions follows that
given in \cite{Abra72}). For given values $u$ and $\mu $ this function has
an infinity of real zeros. We will denote the positive zeros arranged in
ascending order of magnitude as $z_{k}$:%
\begin{equation}
P_{iz_{k}-1/2}^{\mu }(u)=0,\;k=1,2,\ldots .  \label{Pzk0}
\end{equation}%
These zeros are functions of the parameters $u$ and $\mu $: $%
z_{k}=z_{k}(u,\mu )$. Note that one has $P_{iz-1/2}^{\mu
}(u)=P_{-iz-1/2}^{\mu }(u)$ and, hence, $-z_{k}$ are zeros of the function $%
P_{iz-1/2}^{\mu }(u)$ as well. In appendix \ref{sec:Zeros} we show that the
zeros $z_{k}$ are simple and under the conditions specified above the
function $P_{iz-1/2}^{\mu }(u)$\ has no zeros which are not real.

A summation formula for the series over $z_{k}$ can be obtained by making
use of the generalized Abel-Plana formula \cite{Sah1} (see also, \cite%
{Saha07Rev}):%
\begin{equation}
\lim_{b\rightarrow \infty }\left\{ {\mathrm{p.v.}}\!\int_{a}^{b}dx%
\,f(x)-R[f(z),g(z)]\right\} =\frac{1}{2}\int_{a-i\infty }^{a+i\infty }dz\,%
\left[ g(z)+{\sigma (z)}f(z)\right] ,  \label{GAPF}
\end{equation}%
where ${\sigma (z)\equiv \mathrm{sgn}}({{\mathrm{Im\,}}}z)$, the functions $%
f(z)$ and $g(z)$ are meromorphic for $a\leqslant x\leqslant b$ in the
complex plane $z=x+iy$ and p.v. stands for the principal value of the
integral. In formula (\ref{GAPF}) we have defined
\begin{equation}
R[f(z),g(z)]=\pi i\bigg[\sum_{k}\underset{z=z_{g,k}}{\mathrm{Res}}%
g(z)+\sum_{k,{{\mathrm{Im\,}}}z_{f,k}\neq 0}\sigma (z_{f,k})\underset{z={%
\mathrm{\,}}z_{f,k}}{\mathrm{Res}}f(z)\bigg],  \label{Rfg}
\end{equation}%
with $z_{f,k}$ and $z_{g,k}$ being the positions of the poles of the
functions $f(z)$ and $g(z)$ in the strip $a<x<b$.

The functions $f(z)$ and $g(z)$ in formula (\ref{GAPF}) we choose in the form%
\begin{eqnarray}
f(z) &=&\sinh (\pi z)h(z),  \notag \\
g(z) &=&\frac{e^{-i\mu \pi }h(z)}{\pi iP_{iz-1/2}^{\mu }(u)}\left\{ \cos
[\pi (\mu +iz)]Q_{iz-1/2}^{\mu }(u)+\cos [\pi (\mu -iz)]Q_{-iz-1/2}^{\mu
}(u)\right\} ,  \label{gz}
\end{eqnarray}%
where $Q_{iz-1/2}^{\mu }(u)$ is the associated Legendre function of the
second kind and the function $h(z)$ is meromorphic for $a\leqslant {{\mathrm{%
Re}}}\,z\leqslant b$. By using relation (\ref{RelPQ}) between the associated
Legendre functions given in appendix \ref{sec:LegAsymp}, for the combination
appearing on the left hand-side of formula (\ref{GAPF}) one finds%
\begin{equation}
g(z)\pm f(z)=\frac{2e^{-i\mu \pi }h(z)}{\pi iP_{iz-1/2}^{\mu }(u)}\cos [\pi
(\mu \mp iz)]Q_{\mp iz-1/2}^{\mu }(u).  \label{gzplmin}
\end{equation}%
With the functions (\ref{gz}) the expression for $R[f(z),g(z)]$ takes the
form%
\begin{equation}
R[f(z),g(z)]=2\sum_{k}\frac{e^{-i\mu \pi }Q_{iz-1/2}^{\mu }(u)}{\partial
_{z}P_{iz-1/2}^{\mu }(u)}\cos [\pi (\mu +iz)]h(z)\bigg|_{z=z_{k}}+2e^{-i\mu
\pi }r[h(z)],  \label{Rfg2}
\end{equation}%
with the notation%
\begin{eqnarray}
r[h(z)] &=&\sum_{k,{{\mathrm{Im\,}}}z_{h,k}\neq 0}\underset{z=z_{h,k}}{%
\mathrm{Res}}\bigg\{\frac{Q_{-\sigma (z)iz-1/2}^{\mu }(u)}{P_{iz-1/2}^{\mu
}(u)}\cos [\pi (\mu -\sigma (z)iz)]h(z)\bigg\}  \notag \\
&&+\frac{1}{2}\sum_{k,{{\mathrm{Im\,}}}z_{h,k}=0}\underset{z=z_{h,k}}{%
\mathrm{Res}}\bigg\{\frac{h(z)}{P_{iz-1/2}^{\mu }(u)}\sum_{l=\pm }\cos [\pi
(\mu +liz)]Q_{liz-1/2}^{\mu }(u)\bigg\}.  \label{rhz}
\end{eqnarray}%
In formula (\ref{rhz}), $z_{h,k}$ are the positions of the poles for the
function $h(z)$.

In terms of the function $h(z)$ the conditions for the generalized
Abel-Plana formula (\ref{GAPF}) to be valid take the form%
\begin{eqnarray}
\lim_{w\rightarrow \infty }\int_{a\pm iw}^{b\pm iw}dz\frac{Q_{\mp i
z-1/2}^{\mu }(u)}{P_{iz-1/2}^{\mu }(u)}\cos [\pi (\mu \mp i z)]h(z) &=&0,
\notag \\
\lim_{b\rightarrow \infty }\int_{b}^{b\pm i\infty }dz\frac{Q_{\mp i
z-1/2}^{\mu }(u)}{P_{iz-1/2}^{\mu }(u)}\cos [\pi (\mu \mp iz)]h(z) &=&0.
\label{Cond1}
\end{eqnarray}%
By using the asymptotic formulae for the associated Legendre functions given
in appendix \ref{sec:LegAsymp}, it can be seen that these conditions are
satisfied if the function $h(z)$ is restricted to the constraint%
\begin{equation}
|h(z)|<\varepsilon (x)e^{c\eta y},\;z=x+iy,\;|z|\rightarrow \infty ,
\label{Cond2}
\end{equation}%
uniformly in any finite interval of \ $x$, where $c<2$, $\varepsilon
(x)e^{\pi x}\rightarrow 0$ for $x\rightarrow +\infty $, and $\eta $ is
defined by the relation%
\begin{equation}
u=\cosh \eta .  \label{ueta}
\end{equation}

Substituting the functions (\ref{gz}) into formula (\ref{GAPF}) and by
taking into account relations (\ref{gzplmin}), (\ref{Rfg2}), we obtain that
for a function $h(z)$ meromorphic in the half-plane ${{\mathrm{Re}}}%
\,z\geqslant a$ and satisfying condition (\ref{Cond2}), the following
formula takes place%
\begin{eqnarray}
&&\lim_{b\rightarrow \infty }\bigg\{\sum_{k=m}^{n}T_{\mu }(z_{k},u)h(z_{k})-%
\frac{e^{i\mu \pi }}{2}{\mathrm{p.v.}}\!\int_{a}^{b}dx\,\sinh (\pi
x)h(x)+r[h(z)]\bigg\}  \notag \\
&&\qquad =\frac{i}{2\pi }\int_{a-i\infty }^{a+i\infty }dz\,\frac{Q_{-\sigma
(z)iz-1/2}^{\mu }(u)}{P_{iz-1/2}^{\mu }(u)}\cos [\pi (\mu -\sigma
(z)iz)]h(z),  \label{SumForm0}
\end{eqnarray}%
where and in what follows the notation%
\begin{equation}
T_{\mu }(z,u)=\frac{Q_{iz-1/2}^{\mu }(u)}{\partial _{z}P_{iz-1/2}^{\mu }(u)}%
\cos [\pi (\mu +iz)]  \label{Tmu}
\end{equation}%
is used. On the left-hand side of formula (\ref{SumForm0}), $z_{m-1}<a<z_{m}$%
, $z_{n}<b<z_{n+1}$ and in the definition of $r[h(z)]$ the summation goes
over the poles $z_{h,k}$ in the strip $a<{{\mathrm{Re}}}\,z<b$.

Note that from the Wronskian relation for the associated Legendre functions
one has%
\begin{equation}
Q_{iz-1/2}^{\mu }(u)=\frac{e^{i\mu \pi }\Gamma (iz+\mu +1/2)}{%
(u^{2}-1)\Gamma (iz-\mu +1/2)\partial _{u}P_{iz-1/2}^{\mu }(u)},\;z=z_{k}.
\label{QWrons}
\end{equation}%
Now, by taking into account the formula%
\begin{equation}
\frac{\Gamma (iz+\mu +1/2)}{\Gamma (iz-\mu +1/2)}=\pi \frac{|\Gamma (iz-\mu
+1/2)|^{-2}}{\cos [\pi (\mu +iz)]},  \label{GammaRel}
\end{equation}%
for the gamma function, the factor $T_{\mu }(z_{k},u)$ in (\ref{SumForm0})
can also be written in the form
\begin{equation*}
T_{\mu }(z_{k},u)=\frac{\pi e^{i\mu \pi }|\Gamma (iz-\mu +1/2)|^{-2}}{%
(u^{2}-1)\partial _{u}P_{iz-1/2}^{\mu }(u)\partial _{z}P_{iz-1/2}^{\mu }(u)}%
\bigg|_{z=z_{k}}.
\end{equation*}

Taking the limit $a\rightarrow 0$, from (\ref{SumForm0}) one obtains that
for a function $h(z)$ meromorphic in the half-plane ${{\mathrm{Re}}}%
\,z\geqslant 0$ and satisfying the condition (\ref{Cond2}) the following
formula takes place
\begin{eqnarray}
\sum_{k=1}^{\infty }T_{\mu }(z_{k},u)h(z_{k}) &=&\frac{e^{i\mu \pi }}{2}%
\mathrm{p.v.}\int_{0}^{\infty }dx\,\sinh (\pi x)h(x)-r[h(z)]  \notag \\
&&-\frac{1}{2\pi }\int_{0}^{\infty }dx\,\frac{Q_{x-1/2}^{\mu }(u)}{%
P_{x-1/2}^{\mu }(u)}\cos [\pi (\mu +x)][h(xe^{\pi i/2})+h(xe^{-\pi i/2})].
\label{SumFormula}
\end{eqnarray}%
If the function $h(z)$ has poles on the positive real axis, it is assumed
that the first integral on the right-hand side converges in the sense of the
principal value. From the derivation of (\ref{SumFormula}) it follows that
this formula may be extended to the case of some functions $h(z)$ having
branch-points on the imaginary axis, for example, having the form $%
h(z)=h_{1}(z)/(z^{2}+c^{2})^{1/2}$, where $h_{1}(z)$ is a meromorphic
function. This type of function appears in the physical example discussed in
section \ref{sec:Phys}. Special cases of formula (\ref{SumFormula}) with
examples are considered in the next section.

Formula (\ref{SumFormula}) can be generalized for a class of functions $h(z)$
having purely imaginary poles at the points $z=\pm iy_{k}$, $y_{k}>0$, $%
k=1,2,\ldots $, and at the origin $z=y_{0}=0$. Let function $h(z)$ satisfy
the condition%
\begin{equation}
h(z)=-h(ze^{-\pi i})+o((z-\sigma _{k})^{-1}),\;z\rightarrow \sigma
_{k},\;\sigma _{k}=0,iy_{k}.  \label{Impolecond}
\end{equation}%
Now, in the limit $a\rightarrow 0$ the right hand side of (\ref{SumForm0})
can be presented in the form%
\begin{equation}
\frac{i}{2\pi }\sum_{\alpha =+,-}\bigg(\int_{\gamma _{\rho }^{\alpha
}}dz+\sum_{\sigma _{k}=\alpha iy_{k}}\int_{C_{\rho }(\sigma _{k})}dz\bigg)\,%
\frac{Q_{-\alpha iz-1/2}^{\mu }(u)}{P_{iz-1/2}^{\mu }(u)}\cos [\pi (\mu
-\alpha iz)]h(z),  \label{Impoles1}
\end{equation}%
plus the sum of the integrals along the straight segments $(\pm
i(y_{k-1}+\rho ),\pm i(y_{k}-\rho ))$\ of the imaginary axis between the
poles. In (\ref{Impoles1}), $C_{\rho }(\sigma _{k})$ denotes the right half
of the circle with radius $\rho $ and with the center at the point $\sigma
_{k}$, described in the positive direction. Similarly, $\gamma _{\rho }^{+}$
and $\gamma _{\rho }^{-}$ are upper and lower halves of the semicircle in
the right half-plane with radius $\rho $ and with the center at the point $%
z=0$, described in the positive direction with respect to this point. In the
limit $\rho \rightarrow 0$ the sum of the integrals along the straight
segments of the imaginary axis gives the principal value of the last
integral on the right-hand side of (\ref{SumFormula}). Further, in the terms
of (\ref{Impoles1}) with $\alpha =-$ we introduce a new integration variable
$z^{\prime }=ze^{\pi i}$. By using the relation (\ref{Impolecond}) the
expression (\ref{Impoles1}) is presented in the form%
\begin{equation}
-\sum_{\sigma _{k}=0,iy_{k}}(1-\delta _{0\sigma _{k}}/2)\underset{z=\sigma
_{k}}{\mathrm{Res}}\bigg\{\frac{Q_{-iz-1/2}^{\mu }(u)}{P_{iz-1/2}^{\mu }(u)}%
\cos [\pi (\mu -iz)]h(z)\bigg\}  \label{Impoles2}
\end{equation}%
plus the part which vanishes in the limit $\rho \rightarrow 0$. As a result,
formula (\ref{SumFormula}) is extended for functions having purely imaginary
poles and satisfying condition (\ref{Impolecond}). For this, on the
right-hand side of (\ref{SumFormula}) we have to add the sum of residues (%
\ref{Impoles2}) at these poles and take the principal value of the second
integral on the right-hand side. The latter exists due to condition (\ref%
{Impolecond}). Note that for functions having the form $%
h(z)=F(z)P_{iz-1/2}^{\mu }(u)$ the left-hand side of (\ref{SumFormula}) is
zero and from this formula we obtain a formula relating the integrals
involving the Legendre associated functions.

\section{Special cases}

\label{sec:Special}

Here we will consider special cases of the summation formula (\ref%
{SumFormula}). First let us consider the case $\mu =-1/2$. The corresponding
associated Legendre functions have the form%
\begin{equation}
P_{z-1/2}^{-1/2}(\cosh \eta )=\sqrt{\frac{2}{\pi }}\frac{\sinh (z\eta )}{z%
\sqrt{\sinh \eta }},\;Q_{z-1/2}^{-1/2}(\cosh \eta )=-i\sqrt{\frac{\pi }{2}}%
\frac{e^{-z\eta }}{z\sqrt{\sinh \eta }}.  \label{SpCase1}
\end{equation}%
In this case one has $z_{k}=\pi k/\eta $. Introducing a new function $F(x)$
in accordance with the relation $F(\eta x/\pi )=\sinh (\pi x)h(x)$, and
assuming that this function is analytic in the right half-plane, from
formula (\ref{SumFormula}) we find the Abel-Plana formula in the standard
form:%
\begin{equation}
\sum_{k=1}^{\infty }F(k)=-\frac{1}{2}F(0)+\int_{0}^{\infty
}dx\,F(x)+i\int_{0}^{\infty }dx\frac{F(ix)-F(-ix)}{e^{2\pi x}-1}.
\label{AP1}
\end{equation}%
Note that the first term on the right-hand side of this formula comes from
the residue term with $\sigma _{k}=0$ in (\ref{Impoles2}).

In the case $\mu =1/2$ for the corresponding associated Legendre functions
we have the expressions%
\begin{equation}
P_{z-1/2}^{1/2}(\cosh \eta )=\sqrt{\frac{2}{\pi }}\frac{\cosh (z\eta )}{%
\sqrt{\sinh \eta }},\;Q_{z-1/2}^{1/2}(\cosh \eta )=i\sqrt{\frac{\pi }{2}}%
\frac{e^{-z\eta }}{\sqrt{\sinh \eta }}.  \label{SpCase2}
\end{equation}%
The zeros $z_{k}$ now have the form $z_{k}=\pi (k+1/2)/\eta $ and for
functions $F(z)$ analytic in the right half-plane from formula (\ref%
{SumFormula}) we obtain the Abel-Plana formula in the form useful for
fermionic field calculations (see, for instance, \cite{Grib94,Most97}):%
\begin{equation}
\sum_{k=1}^{\infty }F(k+1/2)=\int_{0}^{\infty }dx\,F(x)-i\int_{0}^{\infty }dx%
\frac{F(ix)-F(-ix)}{e^{2\pi x}+1}.  \label{AP2}
\end{equation}

As a next special case let us consider the formula for the summation over
the zeros of the function $P_{isz-1/2}^{-\mu }(\cosh (\eta /s))$ in the
limit when $s\rightarrow \infty $. By taking into account the relation (see
appendix \ref{sec:LegAsymp})%
\begin{equation}
\lim_{\nu \rightarrow +\infty }\nu ^{\mu }P_{i\nu -1/2}^{-\mu }(\cosh (\eta
/\nu ))=J_{\mu }(\eta ),  \label{Plim}
\end{equation}%
with $J_{\mu }(\eta )$ being the Bessel function of the first kind, in this
limit from (\ref{SumFormula}) we obtain the summation formula for the series
over zeros $\eta =j_{\mu ,k}$, $k=1,2,\ldots ,$ of the Bessel function. In
order to take this limit we also will need the formulae (\ref{Qasymp}) from
appendix \ref{sec:LegAsymp} and the formulae \cite{Erde53b}%
\begin{eqnarray}
\lim_{\nu \rightarrow \infty }\nu ^{\mu }P_{\nu }^{-\mu }[\cosh (x/\nu )]
&=&I_{\mu }(x),  \notag \\
\lim_{\nu \rightarrow \infty }\nu ^{\mu }Q_{\nu }^{-\mu }[\cosh (x/\nu )]
&=&e^{-i\mu \pi }K_{\mu }(x),  \label{PQlim}
\end{eqnarray}%
with $I_{\mu }(x)$, $K_{\mu }(x)$\ being the modified Bessel functions.
First we rewrite formula (\ref{SumFormula}) making the replacements $%
z\rightarrow sz$, $x\rightarrow sx$, $\mu \rightarrow -\mu $, in both sides
of this formula including the terms in $r[h(z)]$, and we take $u=\cosh (\eta
/s)$. In order to take the limit $s\rightarrow \infty $ for the second
integral on the right-hand side of the resulting formula, we note that, as
it follows from the derivation of (\ref{SumFormula}), the integrand of this
integral (with the replacements described above) should be understood as the
limit
\begin{equation}
\cos [\pi (sx-\mu )]\sum_{l=+,-}h(sxe^{l\pi i/2})=\lim_{\epsilon \rightarrow
+0}\sum_{l=+,-}\cos [\pi (sx-\mu -lis\epsilon )]h(sxe^{l\pi i/2}).
\label{note1}
\end{equation}%
Taking the limit $s\rightarrow \infty $ with the help of formulae (\ref{Plim}%
),(\ref{PQlim}),(\ref{Qasymp}), we find the following summation formula over
the zeros of the Bessel function%
\begin{eqnarray}
\sum_{k=1}^{\infty }\frac{2f(j_{\mu ,k})}{j_{\mu ,k}J_{\mu }^{\prime
2}(j_{\mu ,k})} &=&\mathrm{p.v.}\int_{0}^{\infty }dx\,f(x)-r_{1}[f(z)]
\notag \\
&&-\frac{1}{\pi }\int_{0}^{\infty }dx\,\frac{K_{\mu }(x)}{I_{\mu }(x)}\left[
e^{i\pi \mu }f(xe^{\pi i/2})+e^{-i\pi \mu }f(xe^{-\pi i/2})\right] ,
\label{SumBess}
\end{eqnarray}%
where $f(z)=\lim_{s\rightarrow \infty }e^{sz/\eta }h(sz/\eta )$, and
\begin{eqnarray}
r_{J}[f(z)] &=&\pi i\sum_{k}\underset{{{\mathrm{Im\,}}}z_{h,k}>0}{\mathrm{Res%
}}\bigg[\frac{H_{\mu }^{(1)}(z)}{J_{\mu }(z)}f(z)\bigg]-\pi i\sum_{k}%
\underset{{{\mathrm{Im\,}}}z_{h,k}<0}{\mathrm{Res}}\bigg[\frac{H_{\mu
}^{(2)}(z)}{J_{\mu }(z)}f(z)\bigg]  \notag \\
&&-\pi \sum_{k}\underset{{{\mathrm{Im\,}}}z_{h,k}=0}{\mathrm{Res}}\bigg[%
\frac{Y_{\mu }(z)}{J_{\mu }(z)}f(z)\bigg]-\frac{\pi }{2}\underset{z=0}{%
\mathrm{Res}}\bigg[\frac{Y_{\mu }(z)}{J_{\mu }(z)}f(z)\bigg].  \label{r1fz}
\end{eqnarray}%
This formula is a special case of the result derived in \cite{Sah1,Sahdis}
(see also, \cite{Saha07Rev}).

Now let us consider two important special cases of (\ref{SumFormula})
corresponding to $\mu =-l$, $h(z)=H(z)/\cosh (\pi z)$ and $\mu =-l-1/2$, $%
h(z)=H(z)/\sinh (\pi z)$ with $l=0,1,2,\ldots $. The associated Legendre
functions with these values of the order appear as radial solutions of the
equations for various fields on background of constant curvature spaces in
cylindrical and spherical coordinates. Let the function $H(z)$ is
meromorphic in the half-plane ${{\mathrm{Re}}}\,z\geqslant 0$ and satisfy
the condition
\begin{equation}
|H(z)|<\varepsilon _{H}(x)e^{c\eta y},\;z=x+iy,\;|z|\rightarrow \infty ,
\label{condhc}
\end{equation}%
uniformly in any finite interval of \ $x>0$, where $c<2$, $\varepsilon
_{H}(x)\rightarrow 0$ for $x\rightarrow +\infty $. Then from the results of
section \ref{sec:SumForm} it follows that the formula
\begin{eqnarray}
\sum_{k=1}^{\infty }\frac{(-1)^{\delta }Q_{iz_{k}-1/2}^{-l-\delta
/2}(u)H(z_{k})}{\partial _{z}P_{iz-1/2}^{-l-\delta /2}(u)|_{z=z_{k}}} &=&%
\frac{1}{2}\mathrm{p.v.}\int_{0}^{\infty }dx\,\tanh ^{1-\delta }(\pi
x)H(x)-r_{\delta }[H(z)]  \notag \\
&-&\frac{1}{2\pi }\int_{0}^{\infty }dx\,\frac{Q_{x-1/2}^{-l-\delta /2}(u)}{%
P_{x-1/2}^{-l-\delta /2}(u)}[H(xe^{\pi i/2})+(-1)^{\delta }H(xe^{-\pi i/2})],
\label{SumFormlcs}
\end{eqnarray}%
takes place, where $\delta =0,1$. In this formula we have introduced the
notation%
\begin{eqnarray}
r_{\delta }[H(z)] &=&\sum_{k,{{\mathrm{Im\,}}}z_{h,k}\neq 0}\underset{%
z=z_{h,k}}{\mathrm{Res}}\bigg[\sigma ^{\delta }(z)\frac{Q_{-\sigma
(z)iz-1/2}^{-l-\delta /2}(u)}{P_{iz-1/2}^{-l-\delta /2}(u)}H(z)\bigg]  \notag
\\
&&+\sum_{k,{{\mathrm{Im\,}}}z_{h,k}=0}\underset{z=z_{h,k}}{\mathrm{Res}}%
\bigg[\frac{Q_{-iz-1/2}^{-l-\delta /2}(u)+(-1)^{\delta
}Q_{iz-1/2}^{-l-\delta /2}(u)}{2P_{iz-1/2}^{-l-\delta /2}(u)}H(z)\bigg]\,.
\label{rdelta}
\end{eqnarray}%
Adding to the right-hand side of formula (\ref{SumFormlcs}) the term
\begin{equation}
-(-1)^{\delta }\sum_{\sigma _{k}=0,iy_{k}}(1-\delta _{0\sigma _{k}}/2)%
\underset{z=\sigma _{k}}{\mathrm{Res}}\bigg[\frac{Q_{-iz-1/2}^{-l-\delta
/2}(u)}{P_{iz-1/2}^{-l-\delta /2}(u)}H(z)\bigg]\,,  \label{ImagPolesls}
\end{equation}%
with $H(z)$ obeying the condition $H(z)=-(-1)^{\delta }H(ze^{-\pi
i})+o((z-\sigma _{k})^{-1})$ for$\;z\rightarrow \sigma _{k}$, we obtain the
extension of this formula to the case when the function $H(z)$ has poles at
the points $0$, $\pm y_{k}$.

From (\ref{SumFormlcs}), as an example when the series is summarized in
closed form one has%
\begin{eqnarray}
\sum_{k=1}^{\infty }\frac{Q_{iz-1/2}^{-l-1/2}(u)}{\partial
_{z}P_{iz-1/2}^{-l-1/2}(u)}\frac{z^{2n}\cos (\alpha z)}{(z^{2}+c^{2})^{m+1}}%
\bigg|_{z=z_{k}} &=&\frac{(-1)^{m+n}}{m!}\bigg\{-\frac{\pi }{2^{m+2}}\left(
\frac{\partial }{c\partial c}\right) ^{m}(c^{2n-1}e^{-\alpha c})  \notag \\
&&-i\frac{\partial ^{m}}{\partial x^{m}}\bigg[\frac{Q_{x-1/2}^{-l-1/2}(u)}{%
P_{x-1/2}^{-l-1/2}(u)}\frac{x^{2n}\cosh (\alpha x)}{(x+c)^{m+1}}\bigg]_{x=c}%
\bigg\}\,,  \label{Example1}
\end{eqnarray}%
where $\alpha <2\eta $,$\;c>0$, with $m\geqslant 0$ and $0\leqslant
n\leqslant m$ being integers. The last term on the right-hand side of this
formula comes from the residue at the pole $\sigma _{k}=ic$. As a next
example, we take in formula (\ref{SumFormlcs}) with $\delta =1$ the function%
\begin{equation}
H(z)=z^{2n-\nu }J_{\nu }(az)\frac{J_{\alpha }(b\sqrt{z^{2}+c^{2}})}{%
(z^{2}+c^{2})^{\alpha /2}},  \label{Example2}
\end{equation}%
where $a$, $b$, $c$ are positive constants and $n$ is a non-negative
integer. This function is analytic in the right half-plane and satisfies the
condition (\ref{condhc}) if $a+b<2\eta $, $2n<\alpha +\nu $. By taking into
account that (\ref{Example2}) is an even function of $z$, from (\ref%
{SumFormlcs}) we find%
\begin{equation}
\sum_{k=1}^{\infty }J_{v}(az_{k})\frac{J_{\alpha }(b\sqrt{z_{k}^{2}+c^{2}})}{%
(z_{k}^{2}+c^{2})^{\alpha /2}}\frac{z_{k}^{\nu +2n}Q_{iz_{k}-1/2}^{-l-1/2}(u)%
}{\partial _{z}P_{iz-1/2}^{-l-1/2}(u)|_{z=z_{k}}}=-\frac{1}{2}%
\int_{0}^{\infty }dx\,x^{2n-\nu }J_{\nu }(ax)\frac{J_{\alpha }(b\sqrt{%
x^{2}+c^{2}})}{(x^{2}+c^{2})^{\alpha /2}}.  \label{Example2n}
\end{equation}

\section{Wightman function inside a spherical boundary in a constant
curvature space}

\label{sec:Phys}

In this section we consider a physical application of the summation formula
derived in section \ref{sec:SumForm}. Namely, we will evaluate the positive
frequency Wightman function for a scalar field and the vacuum expectation
value of the field squared inside a spherical shell in a constant negative
curvature space assuming that the field obeys the Dirichlet boundary
condition on the shell (for quantum effects on background of constant
curvature spaces see, for instance, \cite{Grib94,Most97,Birr82} \ and
references therein).

Consider a quantum scalar field $\varphi (x)$ with the curvature coupling
parameter $\xi $ on background of the space with constant negative curvature
described by the line element
\begin{equation}
ds^{2}=dt^{2}-a^{2}\left[ dr^{2}+\sinh ^{2}r(d\theta ^{2}+\sin ^{2}\theta
d\phi ^{2})\right] ,  \label{metric}
\end{equation}%
where $a$ is a constant which is related to the non-zero components of the
Ricci tensor and the Ricci scalar by the relations%
\begin{equation}
R_{1}^{1}=R_{2}^{2}=R_{3}^{3}=-\frac{2}{a^{2}},\;R=-\frac{6}{a^{2}}.
\label{Rii}
\end{equation}%
The field equation has the form%
\begin{equation}
\left( \nabla _{l}\nabla ^{l}+M^{2}+\xi R\right) \varphi (x)=0,
\label{FieldEq}
\end{equation}%
where $M$ is the mass of the field quanta.

We are interested in quantum effects induced by the presence of a spherical
shell with radius $r=r_{0}$, on which the field obeys Dirichlet boundary
condition: $\varphi (x)|_{r=r_{0}}=0$. This boundary condition modifies the
spectrum of the zero-point fluctuations compared with the case of free space
and changes the physical properties of the vacuum. Among the most important
characteristics of the vacuum are the expectation values of quantities
bilinear in the field operator such as the field squared and the
energy-momentum tensor. These expectation values are obtained from two-point
functions in the coincidence limit. As a two-point function here we will
consider the positive frequency Wightman function $W(x,x^{\prime })=\langle
0|\varphi (x)\varphi (x^{\prime })|0\rangle $, where $|0\rangle $ is the
amplitude for the vacuum state. This function also determines the response
of Unruh-De Witt type particle detectors \cite{Birr82}. Expanding the field
operator over the complete set $\{\varphi _{\alpha }(x),\varphi _{\alpha
}^{\ast }(x)\}$ of classical solutions to the field equation satisfying the
boundary condition, the Wightman function is presented as the mode-sum%
\begin{equation}
W(x,x^{\prime })=\sum_{\alpha }\varphi _{\alpha }(x)\varphi _{\alpha }^{\ast
}(x^{\prime }),  \label{WFsum}
\end{equation}%
where $\alpha $ is a set of quantum numbers specifying the solution.

By the symmetry of the problem under consideration, the eigenfunctions for
the scalar field can be presented in the form%
\begin{equation}
\varphi _{\alpha }(x)=Z(r)Y_{lm}(\theta ,\phi )e^{-i\omega t},
\label{eigfunc1}
\end{equation}%
where $Y_{lm}(\theta ,\phi )$ are standard spherical harmonics, $%
l=0,1,2,\ldots $, $-l\leqslant m\leqslant l$. From the field equation (\ref%
{FieldEq}) we obtain the equation for the radial function $Z(r)$:%
\begin{equation}
\frac{1}{\sinh ^{2}r}\frac{d}{dr}\left( \sinh ^{2}r\frac{dZ}{dr}\right) +%
\left[ (\omega ^{2}-m_{\mathrm{eff}}^{2})a^{2}-\frac{l(l+1)}{\sinh ^{2}r}%
\right] Z=0,  \label{Zeq}
\end{equation}%
where we have introduced the effective mass defined by
\begin{equation}
m_{\mathrm{eff}}^{2}=M^{2}-6\xi /a^{2}.  \label{meff}
\end{equation}%
In the region inside the spherical shell the solution of equation (\ref{Zeq}%
), finite at $r=0$, is expressed in terms of the associated Legendre
function of the first kind\ and the eigenfunctions have the form%
\begin{equation}
\varphi _{\alpha }(x)=C_{\alpha }\frac{P_{iz-1/2}^{-l-1/2}(\cosh r)}{\sqrt{%
\sinh r}}Y_{lm}(\theta ,\phi )e^{-i\omega t},  \label{eigfunc2}
\end{equation}%
with the notation%
\begin{equation}
z^{2}=(\omega ^{2}-m_{\mathrm{eff}}^{2})a^{2}-1.  \label{lambda}
\end{equation}

From the boundary condition on the spherical shell we find that the
eigenvalues for $z$ are solutions of the equation%
\begin{equation}
P_{iz-1/2}^{-l-1/2}(\cosh r_{0})=0,  \label{Eigvalues}
\end{equation}%
and, hence, $z=z_{k}$, $k=1,2,\ldots $, in the notations of section \ref%
{sec:SumForm}. The corresponding eigenfrequencies are found to be%
\begin{equation}
\omega _{k}^{2}=\omega ^{2}(z_{k})=(z_{k}^{2}+1-6\xi )/a^{2}+M^{2}.
\label{eigfreq}
\end{equation}%
Hence, the set $\alpha $ of the quantum numbers is specified to $\alpha
=(l,m,k)$.

The coefficient $C_{\alpha }$ in (\ref{eigfunc2}) is determined from the
orthonormalization condition%
\begin{equation}
\int d^{3}x\,\sqrt{|g|}\varphi _{\alpha }(x)\varphi _{\alpha ^{\prime
}}^{\ast }(x)=\frac{\delta _{\alpha \alpha ^{\prime }}}{2\omega },
\label{normcond}
\end{equation}%
where the integration goes over the region inside the spherical shell.
Substituting the eigenfunctions (\ref{eigfunc2}) into (\ref{normcond}), by
taking into account the integration formula (\ref{int3}) and the boundary
condition, one finds%
\begin{equation}
C_{\alpha }^{-2}=a^{3}\frac{\omega (z)}{z}(u_{0}^{2}-1)\partial
_{z}P_{iz-1/2}^{-l-1/2}(u_{0})\partial
_{u}P_{iz-1/2}^{-l-1/2}(u)|_{z=z_{k},u=u_{0}},  \label{Cnorm}
\end{equation}%
where and in the discussion below we use the notations%
\begin{equation}
u=\cosh r,\;u_{0}=\cosh r_{0}.  \label{uu0}
\end{equation}%
By using the Wronskian relation (\ref{QWrons}), the formula for the
normalization coefficient is written as%
\begin{equation}
C_{\alpha }^{2}=\frac{z_{k}\Gamma
(iz_{k}+l+1)Q_{iz_{k}-1/2}^{-l-1/2}(u_{0})e^{i(l+1/2)\pi }}{a^{3}\omega
(z_{k})\Gamma (iz_{k}-l)\partial _{z}P_{iz-1/2}^{-l-1/2}(u_{0})|_{z=z_{k}}}.
\label{Cnorm2}
\end{equation}%
Note that the ratio of the gamma functions in this formula can also be
presented in the form%
\begin{equation}
\frac{\Gamma (iz_{k}+l+1)}{\Gamma (iz_{k}-l)}=|\Gamma (iz_{k}+l+1)|^{2}\frac{%
\cos [\pi (iz-l-1/2)]}{\pi }.  \label{GamRatio}
\end{equation}

Substituting the eigenfunctions into the mode-sum formula (\ref{WFsum}) and
using the addition theorem for the spherical harmonics, for the Wightman
function one finds%
\begin{eqnarray}
W(x,x^{\prime }) &=&\frac{1}{4\pi ^{2}a^{3}}\sum_{l=0}^{\infty }\frac{%
(2l+1)P_{l}(\cos \gamma )}{\sqrt{\sinh r\sinh r^{\prime }}}e^{i(l+1/2)\pi
}\sum_{k=1}^{\infty }z_{k}|\Gamma (iz_{k}+l+1)|^{2}  \notag \\
&&\times T_{-l-1/2}(z_{k},u_{0})P_{iz_{k}-1/2}^{-l-1/2}(\cosh
r)P_{iz_{k}-1/2}^{-l-1/2}(\cosh r^{\prime })\frac{e^{-i\omega (z_{k})\Delta
t}}{\omega (z_{k})},  \label{WF1}
\end{eqnarray}%
where $\Delta t=t-t^{\prime }$ and $T_{\mu }(z,u)$ is defined by relation (%
\ref{Tmu}). In (\ref{WF1}), $P_{l}(\cos \gamma )$ is the Legendre polynomial
and
\begin{equation}
\cos \gamma =\cos \theta \cos \theta ^{\prime }+\sin \theta \sin \theta
^{\prime }\cos (\phi -\phi ^{\prime }).  \label{cosgam}
\end{equation}%
As the expressions for the zeros $z_{k}$ are not explicitly known, formula (%
\ref{WF1}) for the Wightman function is not convenient. In addition, the
terms in the sum are highly oscillatory for large values of quantum numbers.

For the further evaluation of the Wightman function we apply to the series
over $k$ the summation formula (\ref{SumFormula}) taking in this formula%
\begin{equation}
h(z)=z|\Gamma (iz+l+1)|^{2}P_{iz-1/2}^{-l-1/2}(\cosh
r)P_{iz-1/2}^{-l-1/2}(\cosh r^{\prime })\frac{e^{-i\omega (z)\Delta t}}{%
\omega (z)}.  \label{hz}
\end{equation}%
The corresponding conditions are met if $r+r^{\prime }+\Delta t/a<2r_{0}$.
In particular, this is the case in the coincidence limit $t=t^{\prime }$ for
the region under consideration. For the function (\ref{hz}) the part of the
integral on the right-hand side of formula (\ref{SumFormula}) over the
region $(0,x_{M})$ vanishes and for the Wightman function one finds%
\begin{eqnarray}
W(x,x^{\prime }) &=&W_{0}(x,x^{\prime })-\frac{1}{4\pi ^{2}a^{2}}%
\sum_{l=0}^{\infty }\frac{(2l+1)P_{l}(\cos \gamma )}{\sqrt{\sinh r\sinh
r^{\prime }}}e^{i(l+1/2)\pi }\int_{x_{M}}^{\infty }dx\,x\frac{\Gamma (x+l+1)%
}{\Gamma (x-l)}  \notag \\
&&\times \frac{Q_{x-1/2}^{-l-1/2}(u_{0})}{P_{x-1/2}^{-l-1/2}(u_{0})}%
P_{x-1/2}^{-l-1/2}(\cosh r)P_{x-1/2}^{-l-1/2}(\cosh r^{\prime })\frac{\cosh (%
\sqrt{x^{2}-x_{M}^{2}}\Delta t/a)}{\sqrt{x^{2}-x_{M}^{2}}},  \label{WF2}
\end{eqnarray}%
where we have defined%
\begin{equation}
x_{M}=\sqrt{M^{2}a^{2}+1-6\xi }.  \label{xM}
\end{equation}

In formula (\ref{WF2}), the first term on the right-hand side is given by%
\begin{eqnarray}
W_{0}(x,x^{\prime }) &=&\frac{1}{8\pi ^{2}a^{3}}\sum_{l=0}^{\infty }\frac{%
(2l+1)P_{l}(\cos \gamma )}{\sqrt{\sinh r\sinh r^{\prime }}}e^{-i(l+1/2)\pi
}\int_{0}^{\infty }dx\,x\sinh (\pi x)  \notag \\
&&\times |\Gamma (ix+l+1)|^{2}P_{ix-1/2}^{-l-1/2}(\cosh
r)P_{ix-1/2}^{-l-1/2}(\cosh r^{\prime })\frac{e^{-i\omega (x)\Delta t}}{%
\omega (x)}.  \label{WF0}
\end{eqnarray}%
This function does not depend on the sphere radius and is the Wightman
function for a scalar field in background spacetime described by the line
element (\ref{metric}) when boundaries are absent. This can also be seen by
the direct evaluation. Indeed, when boundaries are absent the eigenfunctions
are still given by formula (\ref{eigfunc2}), where now the spectrum for $z$
is continuous. In this case the corresponding part on the right of the
orthonormalization condition (\ref{normcond}) should be understood as the
Dirac delta function. In the case $z=z^{\prime }$ the normalization integral
diverges and, hence, the main contribution comes from large values $r$. By
using the asymptotic formulae for the associated Legendre functions for
large values of the argument, we can see that%
\begin{equation}
\int_{1}^{\infty }du\,P_{iz-1/2}^{-l-1/2}(u)P_{iz^{\prime
}-1/2}^{-l-1/2}(u)=\left\vert \frac{\Gamma (iz)}{\Gamma (l+1+iz)}\right\vert
^{2}\delta (z-z^{\prime }).  \label{NormInt0}
\end{equation}%
By using this result for the normalization coefficient in the case when
boundaries are absent one finds%
\begin{equation}
C_{\alpha }=\frac{1}{\sqrt{2\omega a^{3}}}\left\vert \frac{\Gamma (l+1+iz)}{%
\Gamma (iz)}\right\vert ,  \label{Calfa}
\end{equation}%
and the eigenfunctions have the form (see also, \cite{Grib94,Grib74})%
\begin{equation}
\varphi _{\alpha }(x)=\left\vert \frac{\Gamma (l+1+iz)}{\Gamma (iz)}%
\right\vert \frac{P_{iz-1/2}^{-l-1/2}(\cosh r)}{\sqrt{2\omega a^{3}\sinh r}}%
Y_{lm}(\theta ,\phi )e^{-i\omega t}.  \label{phialfa0}
\end{equation}%
Substituting these eigenfunctions into the mode-sum (\ref{WFsum}), for the
corresponding Wightman function we find the formula which coincides with (%
\ref{WF0}).

The case of a spherical boundary in the Minkowski spacetime is obtained in
the limit $a\rightarrow \infty $, with fixed $ar=R$. In this limit one has $%
x_{M}=aM$. Introducing a new integration variable $y=x/a$, using the
formulae (\ref{PQlim}) and the asymptotic formula for the gamma function for
large values of the argument, we find%
\begin{eqnarray}
W^{\mathrm{(M)}}(x,x^{\prime }) &=&W_{0}^{\mathrm{(M)}}(x,x^{\prime
})-\sum_{l=0}^{\infty }\frac{(2l+1)P_{l}(\cos \gamma )}{4\pi ^{2}\sqrt{%
RR^{\prime }}}\int_{M}^{\infty }dy\,y  \notag \\
&&\times I_{l+1/2}(Ry)I_{l+1/2}(R^{\prime }y)\frac{K_{l+1/2}(R_{0}y)}{%
I_{l+1/2}(R_{0}y)}\frac{\cosh (\sqrt{y^{2}-M^{2}}\Delta t)}{\sqrt{y^{2}-M^{2}%
}}.  \label{WMink}
\end{eqnarray}%
This formula gives the the positive frequency Wightman function
inside a spherical shell with radius $R_{0}$ in the Minkowski bulk
and is a special case of the general formula given in the first
paper of \cite{Saha01} for a scalar field with Robin boundary
conditions in arbitrary number of spatial dimensions.

Having the Wightman function (\ref{WF2}), we can evaluate the vacuum
expectation value of the field squared taking the coincidence limit of the
argument. Of course, this limit is divergent and some renormalization
procedure is necessary. Here the important point is that for points outside
the spherical shell the local geometry is the same as for the case of
without boundaries and, hence, the structure of the divergences is the same
as well. This is also directly seen from formula (\ref{WF2}), where the
second term on the right-hand side is finite in the coincidence limit. Since
in formula (\ref{WF2}) we have already explicitly subtracted the
boundary-free part, the renormalization is reduced to that for the geometry
without boundaries. In this way for the renormalized vacuum expectation
value of the field squared one has%
\begin{eqnarray}
\langle \varphi ^{2}\rangle _{\mathrm{ren}} &=&\langle \varphi ^{2}\rangle
_{0,\mathrm{ren}}-\sum_{l=0}^{\infty }\frac{e^{i(l+1/2)\pi }}{4\pi ^{2}a^{2}}%
\frac{(2l+1)}{\sinh r}\int_{x_{M}}^{\infty }dx\,x  \notag \\
&&\times \frac{\Gamma (x+l+1)}{\Gamma (x-l)}\frac{Q_{x-1/2}^{-l-1/2}(\cosh
r_{0})}{P_{x-1/2}^{-l-1/2}(\cosh r_{0})}\frac{\left[ P_{x-1/2}^{-l-1/2}(%
\cosh r)\right] ^{2}}{\sqrt{x^{2}-x_{M}^{2}}},  \label{phi2}
\end{eqnarray}%
where the first term on the right-hand side is the corresponding quantity in
the constant negative curvature space without boundaries and the second one
is induced by the presence of the spherical shell. For large values $x$, the
integrand in (\ref{phi2}) behaves as $e^{-(r_{0}-r)x}/(2x\sinh r)$ and the
integral is exponentially convergent at the upper limit for strictly
interior points.

For $r\rightarrow 0$ one has $P_{x-1/2}^{-l-1/2}(\cosh r)\approx
(r/2)^{l+1/2}/\Gamma (l+3/2)$, and in the boundary induced part at the
sphere center the $l=0$ term contributes only:%
\begin{equation}
\langle \varphi ^{2}\rangle _{\mathrm{ren}}=\langle \varphi ^{2}\rangle _{0,%
\mathrm{ren}}-\frac{1}{2\pi ^{2}a^{2}}\int_{x_{M}}^{\infty }dx\,\frac{%
x^{2}(x^{2}-x_{M}^{2})^{-1/2}}{e^{2xr_{0}}-1},\quad r=0.  \label{phi2Cent}
\end{equation}%
where we have used formulae (\ref{SpCase1}). Note that for a conformally
coupled field the boundary induced part in (\ref{phi2Cent}) coincides with
the corresponding quantity for the sphere with radius $ar_{0}$ in the
Minkowski bulk.

\section{Conclusion}

\label{sec:Conclus}

The associated Legendre functions are an important class of special
functions that appear in a wide range of problems of mathematical physics.
In the present paper, specifying the functions in the generalized Abel-Plana
formula in the form (\ref{gz}), we have derived summation formula (\ref%
{SumFormula}) for the series over the zeros of the associated Legendre
function $P_{iz-1/2}^{\mu }(u)$ with respect to the degree. This formula is
valid for functions $h(z)$ meromorphic in the right half-plane and obeying
condition (\ref{Cond2}). Using formula (\ref{SumFormula}), the difference
between the sum over the zeros of the associated Legendre function and the
corresponding integral is presented in terms of an integral involving the
Legendre associated functions with real values of the degree plus residue
terms. For a large class of functions $h(z)$ this integral converges
exponentially fast and, in particular, is useful for numerical calculations.
Frequently used two standard forms of the Abel-Plana formula are obtained as
special cases of formula (\ref{SumFormula}) with $\mu =-1/2$ and $\mu =1/2$
and for an analytic function $h(z)$. Applying the summation formula for the
series over the zeros of the function $P_{iz-1/2}^{\mu }(u\cosh (\eta /s))$
and taking the limit $s\rightarrow \infty $ we have obtained formula (\ref%
{SumBess}) for the summation of the series over zeros of the Bessel
function. The latter is a special case of the formula, previously derived in
\cite{Sah1}. Further, we specify the summation formula for two special cases
of the order $\mu =-l$ and $\mu =-l-1/2$ with $l$ being a non-negative
integer and give examples of the application of this formula. The associated
Legendre functions with these values of the order arise as solutions of the
wave equation on background of constant curvature spaces in cylindrical and
spherical coordinates.

In section \ref{sec:Phys} we consider a physical application of the
summation formula. Namely, for a quantum scalar field we evaluate the
positive frequency Wightman function and the vacuum expectation value of the
field squared inside a spherical shell in a constant negative curvature
space assuming that the field obeys the Dirichlet boundary condition on the
shell. In spherical coordinates the radial part of the corresponding
eigenfunctions contains the function $P_{iz-1/2}^{-l-1/2}(\cosh r)$ and the
eigenfrequencies are expressed in terms of the zeros $z_{k}$ by relation (%
\ref{eigfreq}). As a result, the mode-sum for the Wightman function includes
the summation over these zeros. For the evaluation of the corresponding
series we apply summation formula (\ref{SumFormula}) with the function $h(z)$
given by (\ref{hz}). The term with the first integral on the right-hand side
of formula (\ref{SumFormula}) corresponds to the Wightman function for the
constant curvature space without boundaries and the term with the second
integral is induced by the spherical boundary. For points away from the
shell the latter is finite in the coincidence limit and can be directly used
for the evaluation of the boundary induced part in the vacuum expectation
value of the field squared. The latter is given by the second term on the
right-hand side of formula (\ref{phi2}). The renormalization is necessary
for the boundary-free part only and this procedure is the same as that in
quantum field theory without boundaries.

On the physical example considered we have demonstrated the
advantages for the application of the Abel-Plana-type formulae in
the evaluation of the expectation values of local physical
observables in the presence of boundaries. For the summation of
the corresponding mode-sums the explicit form of the eigenmodes is
not necessary and the part corresponding to the boundary-free
space is explicitly extracted. Further, the boundary induced part
is presented in the form of an integral which rapidly converges
and is finite in the coincidence limit for points away from the
boundary. In this way the renormalization procedure for local
physical observables is reduced to that in quantum field theory
without boundaries. Note that methods for the evaluation of global
characteristics of the vacuum, such as total Casimir energy, in
problems where the eigenmodes are given implicitly as zeros of a
given function, are described in references \cite{Eliz94}.

\section*{Acknowledgements}

The work was supported by the Armenian Ministry of Education and Science
Grant No. 119 and by Conselho Nacional de Desenvolvimento Cient\'{\i}fico e
Tecnol\'{o}gico (CNPq, Brazil).

\appendix

\section{On zeros of the function $P_{iz-1/2}^{\protect\mu }(u)$}

\label{sec:Zeros}

In this appendix we show that the zeros $z=z_{k}$ are simple and real. By
making use of the differential equation for the associated Legendre
functions it can be seen that the following integration formula takes place%
\begin{equation}
\int du\,P_{\nu ^{\prime }}^{\mu }(u)P_{\nu }^{\mu }(u)=(1-u^{2})\frac{%
P_{\nu ^{\prime }}^{\mu }(u)\partial _{u}P_{\nu }^{\mu }(u)-P_{\nu }^{\mu
}(u)\partial _{u}P_{\nu ^{\prime }}^{\mu }(u)}{\nu ^{\prime }(\nu ^{\prime
}+1)-\nu (\nu +1)}+\mathrm{const}.  \label{int1}
\end{equation}%
Taking the limit $\nu ^{\prime }\rightarrow \nu $ and applying the Lopital's
rule for the right-hand side, from this formula we find%
\begin{equation}
\int du\,[P_{\nu }^{\mu }(u)]^{2}=(1-u^{2})\frac{[\partial _{\nu }P_{\nu
}^{\mu }(u)]\partial _{u}P_{\nu }^{\mu }(u)-P_{\nu }^{\mu }(u)\partial _{\nu
}\partial _{u}P_{\nu }^{\mu }(u)}{2\nu +1}+\mathrm{const}.  \label{int2}
\end{equation}%
By taking into account the relation $P_{-iz-1/2}^{\mu }(u)=P_{iz-1/2}^{\mu
}(u)$, we see that for real $z$ one has $[P_{iz-1/2}^{\mu
}(u)]^{2}=|P_{iz-1/2}^{\mu }(u)|^{2}$. Hence, from formula (\ref{int2}) we
find%
\begin{equation}
\int_{1}^{u}dv\,|P_{iz-1/2}^{\mu }(v)|^{2}=\frac{u^{2}-1}{2z}\left\{
[\partial _{z}P_{iz-1/2}^{\mu }(u)]\partial _{u}P_{iz-1/2}^{\mu
}(u)-P_{iz-1/2}^{\mu }(u)\partial _{z}\partial _{u}P_{iz-1/2}^{\mu
}(u)\right\} .  \label{int3}
\end{equation}%
Here we have taken into account that for $u\rightarrow 1$ one has $%
P_{iz-1/2}^{\mu }(u)\sim (u-1)^{-\mu }$ and, hence, $\lim_{u\rightarrow
1}P_{iz-1/2}^{\mu }(u)=0$ for $\mu <0$. From formula (\ref{int3}) it follows
that $[\partial _{z}P_{iz-1/2}^{\mu }(u)]_{z=z_{k}}\neq 0$, and, hence, the
zeros $z_{k}$ are simple.

Now let us show that under the conditions $u>1$ and $\mu \leqslant 0$ all
zeros of the function $P_{iz-1/2}^{\mu }(u)$ are real. Suppose that $%
z=\lambda $ is a zero of $P_{iz-1/2}^{\mu }(u)$ which is not real. As the
function $P_{z-1/2}^{\mu }(u)$ has no real zeros (see, for instance, \cite%
{Grad}), $\lambda $ is not a pure imaginary. If $\lambda ^{\ast }$ is the
complex conjugate to $\lambda $, then it is also a zero of $P_{iz-1/2}^{\mu
}(u)$, because $P_{i\lambda ^{\ast }-1/2}^{\mu }(v)=[P_{i\lambda -1/2}^{\mu
}(v)]^{\ast }$. As a result, from formula (\ref{int1}) we find%
\begin{equation}
\int_{1}^{u}dv\,P_{i\lambda ^{\ast }-1/2}^{\mu }(v)P_{i\lambda -1/2}^{\mu
}(v)=0.  \label{int4}
\end{equation}%
We have obtained a contradiction, since the integrand on the left hand-side
is positive. Hence the number $\lambda $ cannot exist and the function $%
P_{iz-1/2}^{\mu }(u)$ has no zeros which are not real.

From the asymptotic formula (\ref{largey}) for the function $P_{iz-1/2}^{\mu
}(u)$ (see appendix \ref{sec:LegAsymp}\ below) we obtain the asymptotic
expression for large zeros:%
\begin{equation}
z_{k}\sim (\pi k-\pi \mu /2-\pi /4)/\eta .  \label{zkAsymp}
\end{equation}%
Note that this result can also be obtained by taking into account that for
large values $z$ from (\ref{Plim}) one has $P_{iz-1/2}^{-\mu }(\cosh (\eta
))\approx z^{-\mu }J_{\mu }(\eta z)$ and using the asymptotic form for the
zeros of the Bessel function \ (see, for instance, \cite{Abra72}).

\section{Asymptotics of the associated Legendre functions}

\label{sec:LegAsymp}

In this appendix we consider asymptotic expressions for the associated
Legendre functions for large values of the degree. As a starting point we
use the formula
\begin{equation}
Q_{z-1/2}^{\mu }(\cosh \eta )=\sqrt{\pi }e^{i\mu \pi }\frac{\Gamma
(1/2+z+\mu )}{\Gamma (1+z)}\frac{(1-e^{-2\eta })^{\mu }}{e^{(z+1/2)\eta }}%
F(1/2+\mu ,1/2+z+\mu ;1+z;e^{-2\eta }).  \label{Qform1}
\end{equation}%
Using the linear transformation formula 15.3.4 from \cite{Abra72} for the
hypergeometric function, the expression for the function $Q_{z-1/2}^{\mu
}(\cosh \eta )$ is presented in the form%
\begin{equation}
Q_{z-1/2}^{\mu }(\cosh \eta )=\sqrt{\pi }e^{i\mu \pi }\frac{\Gamma
(1/2+z+\mu )}{\Gamma (1+z)}\frac{e^{-z\eta }}{\sqrt{2\sinh \eta }}F(1/2+\mu
,1/2-\mu ;1+z;1/(1-e^{2\eta })).  \label{Qform2}
\end{equation}%
Now, by using the result that for large $|c|$ one has $F(a,b;c;z)=1+O(1/|c|)$%
, from (\ref{Qform2}) the asymptotic formula for the function $%
Q_{z-1/2}^{\mu }(\cosh \eta )$ is obtained for large values $|z|$. The
corresponding formula for the function $P_{z-1/2}^{\mu }(\cosh \eta )$ is
obtained by using the relation%
\begin{equation}
\pi e^{i\mu \pi }\sin (\pi z)P_{z-1/2}^{\mu }(\cosh \eta )=\cos [\pi (z-\mu
)]Q_{-z-1/2}^{\mu }(\cosh \eta )-\cos [\pi (z+\mu )]Q_{z-1/2}^{\mu }(\cosh
\eta ).  \label{RelPQ}
\end{equation}%
In this way we obtain the following formulae%
\begin{eqnarray}
P_{z-1/2}^{\mu }(\cosh \eta ) &\sim &\sqrt{\frac{2}{\pi }}\frac{y^{\mu -1/2}%
}{\sqrt{\sinh \eta }}\sin (\eta y-i\eta x+\pi \mu /2+\pi /4),  \notag \\
Q_{z-1/2}^{\mu }(\cosh \eta ) &\sim &\sqrt{\frac{\pi }{2}}\frac{y^{\mu -1/2}%
}{\sqrt{\sinh \eta }}\exp [-\eta x-i(\eta y-\pi \mu /2-\pi /4)],
\label{largey}
\end{eqnarray}%
in the limit $y\rightarrow +\infty $, $z=x+iy$, and the formulae%
\begin{eqnarray}
P_{z-1/2}^{\mu }(\cosh \eta ) &\sim &\frac{x^{\mu -1/2}}{\sqrt{2\pi \sinh
\eta }}e^{\eta x+i\eta y},  \notag \\
Q_{z-1/2}^{\mu }(\cosh \eta ) &\sim &\sqrt{\frac{\pi }{2}}e^{i\mu \pi }\frac{%
x^{\mu -1/2}}{\sqrt{\sinh \eta }}e^{-\eta x-i\eta y},  \label{largex}
\end{eqnarray}%
in the limit $x\rightarrow +\infty $.

Now let us consider the asymptotics of the functions $P_{i\nu -1/2}^{-\mu
}(\cosh (\eta /\nu ))$, $Q_{\pm i\nu -1/2}^{-\mu }(\cosh (\eta /\nu ))$ as $%
\nu \rightarrow +\infty $. These asymptotics are obtained in the way similar
to that used in \cite{Erde53b} for formulae (\ref{PQlim}). Our starting
point is the formula
\begin{equation}
P_{i\nu -1/2}^{-\mu }(\cosh (\eta /\nu ))=\frac{\tanh ^{\mu }(\eta /2\nu )}{%
\Gamma (1+\mu )}F(1/2-i\nu ,1/2+i\nu ;1+\mu ;-\sinh ^{2}(\eta /2\nu )),
\label{PtoF}
\end{equation}%
relating the associated Legendre function to the hypergeometric function.
From the definition of the hypergeometric function it is not difficult to
see that
\begin{equation}
\lim_{\nu \rightarrow +\infty }F(1/2-i\nu ,1/2+i\nu ;1+\mu ;-\sinh ^{2}(\eta
/2\nu ))=\Gamma (1+\mu )(2/\eta )^{\mu }J_{\mu }(\eta ).  \label{Flim}
\end{equation}%
Combining (\ref{PtoF}) and (\ref{Flim}) we obtain formula (\ref{Plim}). The
corresponding formula for the functions $Q_{\pm i\nu -1/2}^{-\mu }(\cosh
(\eta /\nu ))$ are obtained by making use of the relation%
\begin{equation}
\frac{2}{\pi }\sin (\mu \pi )e^{i\mu \pi }Q_{\pm i\nu -1/2}^{-\mu }(u)=\frac{%
\Gamma (\pm i\nu -\mu +1/2)}{\Gamma (\pm i\nu +\mu +1/2)}P_{i\nu -1/2}^{\mu
}(u)-P_{i\nu -1/2}^{-\mu }(u),  \label{QPrel}
\end{equation}%
and formula (\ref{Plim}). In this way we find

\begin{equation}
\lim_{\nu \rightarrow +\infty }\nu ^{\mu }e^{i\mu \pi }Q_{\pm i\nu
-1/2}^{-\mu }(\cosh (\eta /\nu ))=\pi \frac{e^{\mp i\mu \pi }J_{-\mu }(\eta
)-J_{\mu }(\eta )}{2\sin (\mu \pi )},  \label{Qlim}
\end{equation}%
or in the equivalent form
\begin{eqnarray}
\lim_{\nu \rightarrow +\infty }\nu ^{\mu }e^{i\mu \pi }Q_{i\nu -1/2}^{-\mu
}(\cosh (\eta /\nu )) &=&-\frac{\pi i}{2}e^{-i\mu \pi }H_{\mu }^{(2)}(\eta ),
\notag \\
\lim_{\nu \rightarrow +\infty }\nu ^{\mu }e^{i\mu \pi }Q_{-i\nu -1/2}^{-\mu
}(\cosh (\eta /\nu )) &=&\frac{\pi i}{2}e^{i\mu \pi }H_{\mu }^{(1)}(\eta ),
\label{Qasymp}
\end{eqnarray}%
where $H_{\mu }^{(1,2)}(\eta )$ are the Hankel functions.


\begin{thebibliography}{99}
\bibitem{Hard91} G.H. Hardy, \textit{Divergent Series} (Chelsea Publishing
Company, New York, 1991).

\bibitem{Henr74} P. Henrici, \textit{Applied and Computational Complex
Analysis,} Vol. 1 (Wiley, New York, 1974).

\bibitem{Saha07Rev} A.A. Saharian, "The generalized Abel-Plana formula with
applications to Bessel functions and Casimir effect," Preprint
ICTP/2007/082; arXiv: 0708.1187.

\bibitem{Grib94} A.A. Grib, S.G. Mamayev, and V.M. Mostepanenko, \textit{%
Vacuum Quantum Effects in Strong Fields} (Friedmann Laboratory Publishing,
St. Petersburg, 1994).

\bibitem{Most97} V.M. Mostepanenko and N.N. Trunov, \textit{The Casimir
Effect and Its Applications} (Oxford University Press, Oxford, 1997).

\bibitem{Sah1} A.A. Saharian, Izv. AN Arm. SSR. Matematika \textbf{22}, 166
(1987) [Sov. J. Contemp. Math. Analysis, \textbf{22}, 70 (1987)].

\bibitem{Bart80} G. Barton, J. Phys. A \textbf{14}, 1009 (1981); G. Barton,
J. Phys. A \textbf{15}, 323 (1982).

\bibitem{Zaya88} Yu.B Zayaev and I.Yu. Sokolov, In Abstracts of the 3rd
All-Union Conference on "Quantum Metrology and Fundamental Physical
Constants" (VNIIM "D.I. Mendeleev", Leningrad, 1988 (in Russian)).

\bibitem{Sahdis} A.A. Saharian, "Electromagnetic vacuum effects in presence
of macroscopic bodies," PhD thesis (Yerevan, 1987, in Russian).

\bibitem{Saha00Rev} A.A. Saharian, "The generalized Abel-Plana formula.
Applications to Bessel functions and Casimir effect," Preprint IC/2000/14;
hep-th/0002239.

\bibitem{Saha06PoS} A.A. Saharian, Generalized Abel-Plana formula as a
renormalization tool in quantum field theory with boundaries. Proceedings of
the Fifth International Conference on Mathematical Methods in Physics, 24-28
April, 2006, Rio de Janeiro, Brazil; PoS(IC2006)019.

\bibitem{Plun86} G. Plunien, B. Muller, and W.Greiner, Phys. Rept. \textbf{134%
}, 87 (1986); M. Bordag, U. Mohidden, and V.M. Mostepanenko, Phys. Rept.
\textbf{353}, 1 (2001); K.A. Milton, \textit{The Casimir effect: Physical
Manifestation of Zero-Point Energy} (World Scientific, Singapore, 2002).

\bibitem{Rome02} A. Romeo and A.A. Saharian, J. Phys. A \textbf{35}, 1297
(2002); A.A. Saharian and G. Esposito, J. Phys. A \textbf{39}, 5233 (2006).

\bibitem{Saha01} A.A. Saharian, Phys. Rev. D \textbf{63}, 125007 (2001);
A.A. Saharian and M.R. Setare, Class. Quantum Grav. \textbf{20}, 3765
(2003); A.A. Saharian, Astrophys. \textbf{47}, 303 (2004); A.A. Saharian and
M.R. Setare, Int. J. Mod. Phys. A \textbf{19}, 4301 (2004); A.A. Saharian
and E.R. Bezerra de Mello, J. Phys. A \textbf{37}, 3543 (2004); A. A.
Saharian and E. R. Bezerra de Mello, Int. J. Mod. Phys. A \textbf{20}, 2380
(2005); E.R. Bezerra de Mello and A.A. Saharian, Class. Quantum Grav.
\textbf{23}, 4673 (2006).

\bibitem{Rome01} A. Romeo and A.A. Saharian, Phys. Rev. D \textbf{63},
105019 (2001); A.A. Saharian and A.S. Tarloyan, J. Phys. A \textbf{39},
13371 (2006); E.R. Bezerra de Mello, V.B. Bezerra, A.A. Saharian, and A.S.
Tarloyan, Phys. Rev. D \textbf{74}, 025017 (2006); E.R. Bezerra de Mello,
V.B. Bezerra, and A.A. Saharian, Phys. Lett. B \textbf{645}, 245 (2007).

\bibitem{Reza02} A.H. Rezaeian and A.A. Saharian, Class. Quantum Grav.
\textbf{19}, 3625 (2002); A.A. Saharian and A.S. Tarloyan, J. Phys. A
\textbf{38}, 8763 (2005); A.A. Saharian, Eur. Phys. J. C \textbf{52}, 721
(2007); A.A. Saharian and A.S. Tarloyan, Ann. Phys., in press,
arXiv:0708.2970 [hep-th].

\bibitem{SahaRind1} A.A. Saharian, Class. Quantum Grav. \textbf{19}, 5039
(2002); R.M. Avagyan, A.A. Saharian, and A.H. Yeranyan, Phys. Rev. D \textbf{%
66}, 085023 (2002); A.A. Saharian, R.M. Avagyan, and R.S. Davtyan, Int. J.
Mod. Phys. A \textbf{21}, 2353 (2006).

\bibitem{Saha05b} A.A. Saharian, Nucl. Phys. B \textbf{712}, 196 (2005);
A.A. Saharian, Phys. Rev. D \textbf{73}, 044012 (2006); A.A. Saharian, Phys.
Rev. D \textbf{73}, 064019 (2006).

\bibitem{Saha07RindBr} A.A. Saharian and M.R. Setare, JHEP \textbf{0702},
089 (2007).

\bibitem{Saha07Helic} A.A. Saharian, A.S. Kotanjyan, and M.L. Grigoryan, J.
Phys. A \textbf{40}, 1405 (2007).

\bibitem{Erde53a} A. Erd\'{e}lyi \textit{et al}, \textit{Higher
Transcendental Functions}, Vol. 1 (McGraw Hill, New York, 1953).

\bibitem{Abra72} \textit{Handbook of Mathematical Functions}, edited by M.
Abramowitz and I. A. Stegun (Dover, New York, 1972).

\bibitem{Birr82} N.D. Birrell and P.C.W. Davis, \textit{Quantum Fields in
Curved Space} (Cambridge University Press, Cambridge, England, 1982).

\bibitem{Noji00} S. Nojiri, S. Odintsov, and S. Zerbini, Class. Quantum
Grav. \textbf{17}, 4855 (2000); W. Naylor and M. Sasaki, Phys.
Lett. B \textbf{542}, 289 (2002); E. Elizalde, S. Nojiri, S.D.
Odintsov, and S. Ogushi, Phys. Rev. D \textbf{67}, 063515 (2003);
I.G. Moss, W. Naylor, W. Santiago-Germ\'{a}n, M. Sasaki, Phys.
Rev. D \textbf{67}, 125010 (2003); A. Flachi, A. Knapman, W.
Naylor, and M. Sasaki, Phys. Rev. D \textbf{70}, 124011 (2004);
J.P. Norman, Phys.Rev. D \textbf{69}, 125015 (2004); W. Naylor and
M. Sasaki, Prog. Theor. Phys. \textbf{113}, 535 (2005); M.
Minamitsuji, W. Naylor, and M. Sasaki, Nucl. Phys. B \textbf{737},
121 (2006).

\bibitem{Erde53b} A. Erd\'{e}lyi \textit{et al}, \textit{Higher
Transcendental Functions}, Vol. 2 (McGraw Hill, New York, 1953).

\bibitem{Grib74} A.A. Grib, B.A. Levitskii, and V.M. Mostepanenko, Theor.
Math. Phys. \textbf{19}, 349 (1974).

\bibitem{Eliz94} E. Elizalde, S.D. Odintsov, A. Romeo, A.A. Bytsenko, and S.
Zerbini, Zeta regularization techniques with applications (World Scientific,
Singapore, 1994); E. Elizalde, S. Leseduarte, and A. Romeo, J. Phys. A \textbf{26%
}, 2409 (1993); S. Leseduarte and A. Romeo, J. Phys. A
\textbf{27}, 2483 (1994); M. Bordag, J. Phys. A \textbf{28}, 755
(1995); M. Bordag and K. Kirsten, Phys. Rev. D \textbf{53}, 5753
(1996);  M. Bordag, E. Elizalde, and K. Kirsten, J. Math. Phys.
\textbf{37}, 895 (1996); S. Leseduarte and A. Romeo, Ann. Phys.
\textbf{250}, 448 (1996); M. Bordag, K. Kirsten, and J.S. Dowker,
Commun. Math. Phys. \textbf{182}, 371 (1996); M. Bordag, E.
Elizalde, K. Kirsten, and S. Leseduarte, Phys. Rev. D \textbf{56},
4896 (1997); E. Elizalde, M. Bordag, and K. Kirsten, J. Phys. A
\textbf{31}, 1743 (1998); V.V. Nesterenko and I.G. Pirozhenko,
Phys. Rev. D 57, 1284 (1998).

\bibitem{Grad} I.S. Gradshtein and I.M. Ryzhyk, \textit{Tables of Integrals,
Series and Products} (Academic Press, New York, 1980).

\end{thebibliography}
\end{document}